\documentclass[useAMS,usenatbib]{mn2e}

\usepackage{graphicx}

\title[The S2N2 metallicity calibrator and the abundance gradient of M~33]{The S2N2 metallicity calibrator and the abundance gradient of M~33}

\author[K. Viironen et al.]{K. Viironen$^{1}$\thanks{E-mail: kerttu@iac.es (KV);
gloria@inaoep.mx (GDI); amr@iac.es (AM); laura@arcetri.astro.it (LM);
rcorradi@ing.iac.es (RLMC)}, G. Delgado-Inglada$^{2}$,  A. Mampaso$^{1}$,
\newauthor L. Magrini$^{3}$ and R. L. M. Corradi$^{4,1}$ \\
$^{1}$Instituto de Astrof{\'{i}}sica de Canarias (IAC), C/ V{\'{i}}a L\' actea s/n, 38200 La Laguna, Tenerife, Spain\\
$^{2}$Instituto Nacional de Astrof{\'{i}}sica, \'Optica y Electr\' onica  (INAOE), Luis Enrique Erro 1, Tonantzintla, Puebla, Mexico\\
$^{3}$INAF-Osservatorio Astrofisico di Arcetri, Largo E. Fermi, 5 50125, Firenze, Italy\\
$^{4}$Isaac Newton Group of Telescopes, Apartado de Correos 321, E-38700 Sta. Cruz de La Palma, Spain}

\begin{document}

\date{Accepted ?. Received ?; in original form ?}

\pagerange{\pageref{firstpage}--\pageref{lastpage}} \pubyear{2007}

\maketitle

\label{firstpage}

\begin{abstract}
We introduce the
log(H$\alpha$/[S\,\textsc{ii}]$\lambda\lambda$6717+6731)
vs. log(H$\alpha$/[N\,\textsc{ii}]$\lambda$6583) (S2N2) diagnostic
diagram as metallicity and ionisation parameter indicator for
H\,\textsc{ii} regions in external galaxies. The location of
H\,\textsc{ii} regions in the S2N2 diagram was studied both
empirically and theoretically. We found that, for a wide range of
metallicities, the S2N2 diagram gives single valued results in the
metallicity-ionisation parameter plane. We demonstrate that the S2N2
diagram is a powerful tool to estimate metallicities of high-redshift ($z$ $\sim$
2) H\,\textsc{ii} galaxies. Finally, we derive the metallicity for 76
H\,\textsc{ii} regions in M33 from the S2N2 diagram and calculate an
O/H abundance gradient for this galaxy of -0.05 ($\pm$0.01) dex
kpc$^{-1}$.
\end{abstract}

\begin{keywords}
H\,\textsc{ii} regions -- galaxies: abundances -- galaxies: individual: M33 -- galaxies: high-redshift
\end{keywords}

\section{Introduction}

Chemical abundances of the ionised interstellar medium in our own and
in external galaxies are commonly studied by analysing the emission
line spectra of their H\,\textsc{ii} regions. This task is relatively
straightforward if the lines for the electron temperature
determination can be measured directly from the observations. However,
these auroral lines (like [N\,\textsc{ii}]$\lambda$5755 or
[O\,\textsc{iii}]$\lambda$4363) are usually too faint to be detected
for distant H\,\textsc{ii} regions, and especially in high metallicity
regions. This has woken an interest in having abundance indicators
based on strong lines only. The use of these strong-line methods was
pioneered by \citet{pagel79} and \citet{alloin79}, followed later by
numerous studies from different authors. The lines most commonly used
are oxygen ([O\,\textsc{ii}]$\lambda\lambda$3727,3729,
[O\,\textsc{iii}]$\lambda\lambda$4959,5007), nitrogen
([N\,\textsc{ii}]$\lambda$6583), and sulphur
([S\,\textsc{ii}]$\lambda\lambda$6717,6731,
[S\,\textsc{iii}]$\lambda\lambda$6312,9069,9532), normalised to the
hydrogen lines (H$\alpha \lambda$6563, H$\beta \lambda$4861)
\citep{kewley02}. Recently, \citet{stasinska06} presented two new
oxygen abundance indicators based on argon and sulphur lines:
[Ar\,\textsc{iii}]/[O\,\textsc{iii}] and
[S\,\textsc{iii}]/[O\,\textsc{iii}].

\begin{figure*}
  \includegraphics[width=0.5\linewidth,angle=0]{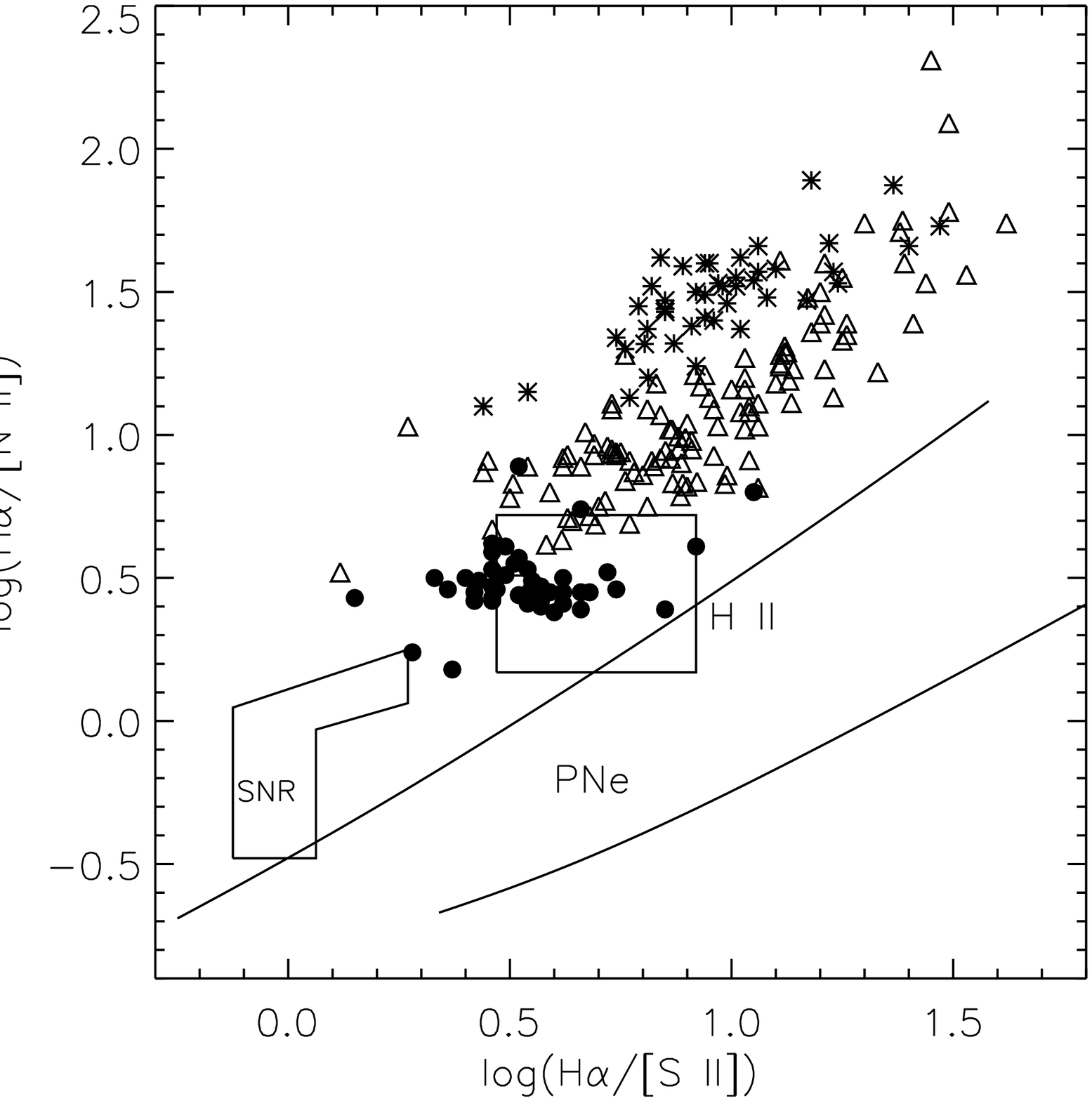}
      \caption{The loci of Galactic H\,{\sc ii} regions, SNRs and PNe
      \citep{sabbadin77} in the S2N2 diagram. The Galactic H\,{\sc ii}
      regions occupy a very limited area in this diagram (the box
      labelled H\,{\sc ii}). Over-plotted is our sample of 205
      extragalactic H\,{\sc ii} regions in Local Group galaxies coded
      according to the metallicity of their parent galaxies, in the
      following way: 7.30 $\le$ 12 + log(O/H) $\le$ 8.06 (stars), 8.06 $<$ 12 + log(O/H) $\le$ 8.40 (triangles), 12 + log(O/H) $>$ 8.40 (filled
      circles).}
\label{boxes}
    \end{figure*}

In this paper we study the
log(H$\alpha$/[S\,\textsc{ii}]$\lambda\lambda$6717+6731)
vs. log(H$\alpha$/[N\,\textsc{ii}]$\lambda$6583) diagram (referred to
as S2N2 in the following) as a metallicity indicator. This diagram was
introduced by \citet{sabbadin77} as a useful tool to separate galactic
planetary nebulae (PNe), H\,\textsc{ii} regions, and SNRs from each others,
and was later applied to Herbig-Haro objects by
\citet{canto81}, to galactic PNe by \citet{garcia_lario91} and
\citet{riesgo05}, and to extragalactic PNe by
\citet{magrini03}. \citet{raimann00} found a correlation between
[N\,\textsc{ii}]/H$\alpha$ and the metallicity for a sample of
H\,\textsc{ii} galaxies. A similar correlation was found by
\citet{van_zee00} for a sample of dwarf galaxies and extra-galactic
H\,\textsc{ii} regions, and was further studied by several authors
\citep[e.g.][]{denicolo02,kewley02,melbourne02,pettini04,maier05,nagao06}. Diagrams
combining [N\,\textsc{ii}]/H$\alpha$ with
[O\,\textsc{iii}]$\lambda$5007/H$\beta$ have been used in object
classification by e.g. \citet{contini02,maier05,groves06}, and
\citet{stasinska06}. The [N\,\textsc{ii}]/H$\alpha$ ratio has been
used to determine ionisation parameter and metallicity in combination
with several line ratios: [O\,\textsc{iii}]$\lambda$5007/H$\beta$
\citep[e.g.][]{dopita00,kewley01}, ([O\,\textsc{ii}]$\lambda$3727 +
[O\,\textsc{iii}]$\lambda\lambda$4959,5007)/H$\beta$ (R$_{23}$)
\citep{melbourne02,lilly03}, and
[O\,\textsc{iii}]$\lambda\lambda$4959,5007/[O\,\textsc{ii}]$\lambda$3727
(O$_{23}$) \citep[e.g.][]{lilly03}. On the other hand,
\citet{denicolo02} demonstrated that the [S\,\textsc{ii}]/H$\alpha$
line ratio is also correlated with the metallicity in a sample of
H\,\textsc{ii} regions and galaxies. The use of a diagram combining
[S\,\textsc{ii}]$\lambda\lambda$6716,6731/H$\alpha$ with
[O\,\textsc{iii}]/H$\beta$ in object classification has been examined
by e.g. \citet{contini02} and \citet{stasinska06}. The use of the same
diagram for abundance and ionisation parameter diagnoses was studied
e.g. by \citet{dopita00}, and
\citet{kewley01}. \citet{perez-montero05} provide an analysis of the
ranges of validity and average uncertainties for several line ratios
used in nebular metallicity studies.

\citet{delgado05} first introduced the use of the S2N2 diagram for
metallicity and ionisation parameter diagnoses for extragalactic
H\,\textsc{ii} regions. \citet{moustakas06} studied whether there was
a difference between integrated spectra of galaxies and the spectra of
individual H\,\textsc{ii} regions when plotted in this
diagram. Recently, \citet{dopita06} used the S2N2 diagram, among many
others, for abundance diagnostics. They calculated time-dependent
photo-ionisation models of individual H\,\textsc{ii} regions and
compared the models with H\,\textsc{ii} region data from different
authors.

In this paper we further explore the use of the S2N2 diagram for
abundance and ionisation parameter diagnoses. We show that the diagram
gives single valued results for a wide range of metallicities. A
comparison of empirical H\,\textsc{ii} region data with
photo-ionisation (\textsc{CLOUDY}) model abundances is made, and a good
fit between data and the models is found up to the highest abundances
expected for H\,\textsc{ii} regions. We also demonstrate that the S2N2
diagram is powerful enough to determine metallicities for
high-redshift (z $\sim$ 2) galaxies using current near-infrared
spectrographs, and sensitive enough to measure with reasonable accuracy
the O/H radial abundance gradient in M33.

\begin{figure*}
   \centering
   \includegraphics[width=0.75\linewidth,angle=0]{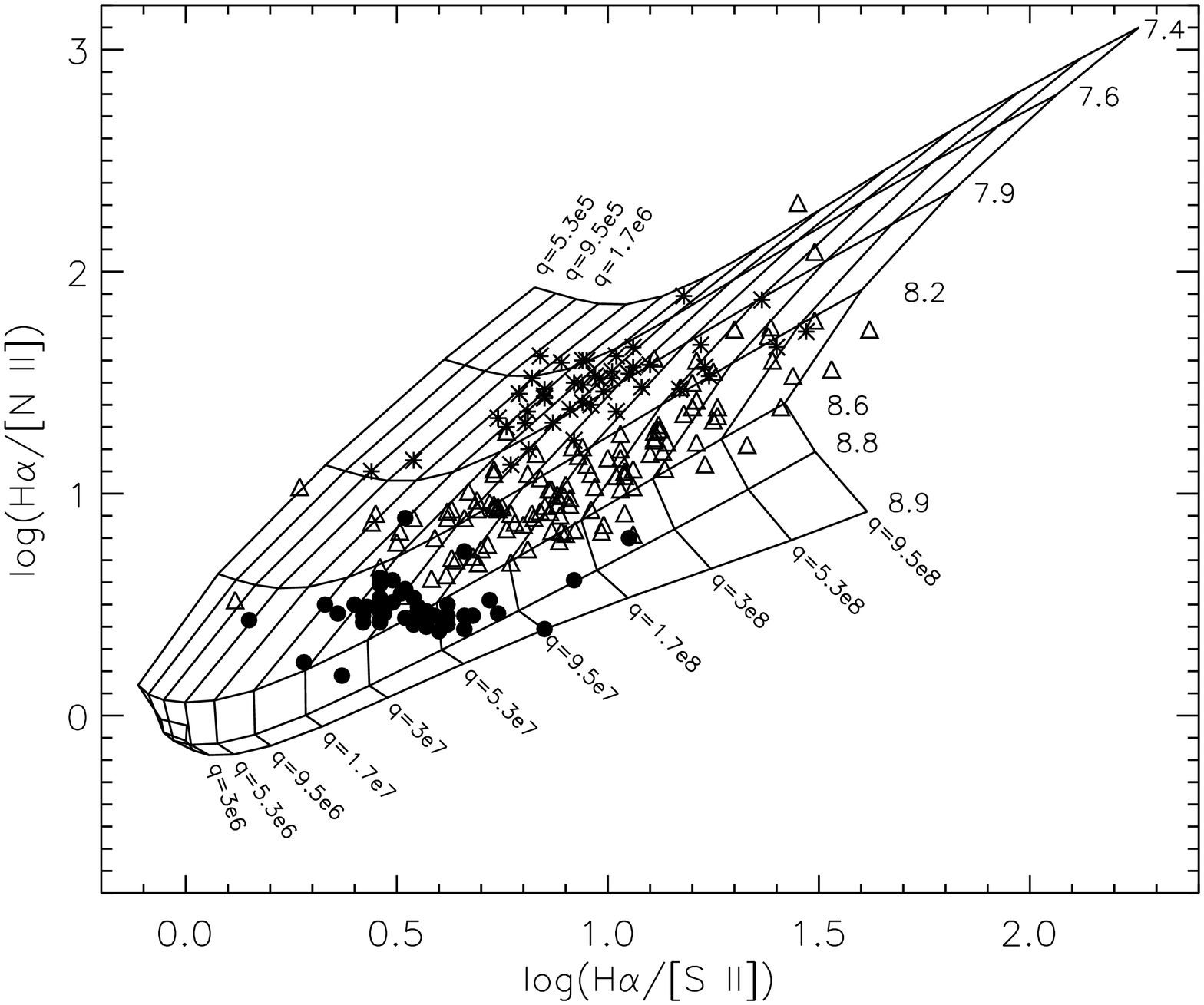}
      \caption{The H\,{\sc ii} region sample as in Fig.~\ref{boxes},
      over-plotted by the grid of photo-ionisation models. Labels
      indicate the values for the ionisation parameter ($5.3\times
      10^5 < q < 9.5\times 10^8$) and oxygen abundance (7.4 $<$
      12+log(O/H) $<$ 8.9) used.}
\label{diagram}
    \end{figure*}

\section[]{Observational data}

\subsection{Sample selection and properties}\label{sec3}

We have collected data for all H\,\textsc{ii} regions in the Local
Group galaxies for which adequate information about the emission lines
needed was available. Data showing errors in the logarithmic line
ratio (log(H$\alpha$/[N\,\textsc{ii}] or
log(H$\alpha$/[S\,\textsc{ii}])) greater than 0.15 dex were
discarded. In the case of H\,\textsc{ii} regions with various
measurements from different authors, the most accurate measurements
were adopted.
Data for the galaxies Sextans A, Sextans B and NGC3109 were also
included, although these galaxies are probably members of a small
association of dwarf galaxies located just at the outskirts of the
Local Group \citep{magrini06}. In addition, the sample includes new data 
for 64 H\,\textsc{ii} regions in M33 published recently by \cite{magrini07}.
 Our final sample consists of 13 galaxies with a
total of 205 well-measured H\,\textsc{ii} regions. The sample
galaxies, their magnitudes, O/H abundances and distances are presented
in order of decreasing O/H abundance in Table \ref{table1}. The data
for the galaxies are adopted from \citet{van_den_bergh00} except for
Gr8 for which the magnitude and O/H abundance measurements are from
\citet{van_zee06}, for EGB0427+63 \citep{hodge95} and for
NGC 3109 \citep{richer95}. The distance measurement for M33 is
from \citet{freedman91}. Data for individual H\,\textsc{ii} regions
and the corresponding references are given in Table \ref{table2}.

Our sample is, in summary, limited to Local Group galaxies, i.e. to
relatively nearby objects, so avoiding biases towards the brightest
H\,\textsc{ii} regions or to H\,\textsc{ii} galaxies. Only spatially
resolved H\,\textsc{ii} regions are included, and the contribution
from stars, clusters, and the diffuse ionised gas along the line of
sight is minimised in the spectra. The sizes of the H\,\textsc{ii}
regions vary from compact to giant ones, and the sample coverage in
luminosity and metallicity is broad: the galaxy magnitudes range from
M$_V$ = -21.2 (M31) to M$_V$ = -10.7 (SagDIG), and the metallicities
vary from 12+log(O/H) = 7.3 (Gr8, Leo A) to 12+log(O/H) = 9.0
(M~31\footnote {this value given in \citet{van_den_bergh00} may be an
overestimation considering the recent studies about the maximum
metallicity in spiral galaxies, i.e. Pilyugin et al. (2007). See Sect. 3.}).
 These can be compared to the less metallic H\,\textsc{ii}
region known, 1Zw18-NW, with a metallicity of 12 + log(O/H) = 7.07
\citep{izotov99}, and to the most metal rich H\,\textsc{ii} regions in
spiral galaxies, with 12 + log(O/H) $\sim$ 8.9 \citep{pilyugin07}.

\subsection{The S2N2 diagram}


The sample H\,\textsc{ii} regions were plotted into the original
\citet{sabbadin77} diagram (Fig.~\ref{boxes}) binned into three
groups, according to the metallicity adopted for their host galaxies: 7.30 
$\le$ 12 + log(O/H) $\le$ 8.06, 8.06 $<$ 12 + log(O/H) $\le$ 8.40, 
and 12 + log(O/H) $>$ 8.40. The error bars
(given in Table \ref{table2}) are not plotted since they are smaller
than, or about the same size as, the symbols. Fig. \ref{boxes} shows
that i) the extragalactic H\,\textsc{ii} regions are not located in
the same region as the galactic ones \citep[in contrast to
extragalactic PNe, which roughly overlap with the galactic
ones,][]{magrini03}, and ii) there is a clear correlation between the
locations of the H\,\textsc{ii} regions in the diagram and their
metallicity, so that as the metallicity increases, both the S2 and N2
line ratios decrease.

\begin{table}
\centering
\caption{The properties of the galaxies used in this study}
\label{table1}
\begin{center}
\begin{tabular}{c c c c c c}
\hline
Galaxy & $M_V$ & 12 + log(O/H) & Distance [kpc]\\
\hline
   M31 &  -21.2 & 9.00 & 760 \\
   M33 &  -18.9 & 8.40 & 840 \\
   LMC &  -18.5 & 8.37 & 50 \\
   IC 10 & -16.3 & 8.20 & 660 \\
   NGC 6822 & -15.96 & 8.14 & 500 \\
   NGC 3109 & -15.80 & 8.06 & 1400 \\
   SMC & -17.07 & 8.02 & 59 \\
   IC 1613 & -15.3 & 7.86 & 725 \\
   WLM & -14.4 & 7.74 & 925 \\
   Sextans B & -14.3 & 7.56 & 1320 \\
   Sextans A & -14.2 & 7.49 & 1450 \\
   Leo A & -11.54 & 7.30 & 690 \\
   Gr8 & -12.5 & 7.30 & 220 \\
\hline
\end{tabular}
\end{center}
\end{table}

\section{Photo-Ionisation modelling}\label{section3}

H\,\textsc{ii} regions were modelled using the photo-ionisation code
\textsc{CLOUDY} C06.02 \citep{ferland98}, assuming a spherical cloud
ionised by a single central source of \citet{kurucz79,kurucz91}
line-blanketed LTE atmosphere with a temperature of 45000 K. The
electron density was assumed to be 100 cm$^{-3}$, a typical value for
giant extragalactic H\,\textsc{ii} regions \citep{stasinska04}. A
filling factor of
0.001 was used, based on a study by \citet{munoz01}, who determined
the filling factor for 73 H\,\textsc{ii} regions in the galaxies NGC
3359 and NGC 7479. For the inner and outer radii of the cloud the
default values of \textsc{CLOUDY} were used implying a thin
bubble-like cloud surrounding the central object and making the
geometry basically plane-parallel. All the chemical elements except
the nitrogen were assumed to be of primary origin, i.e. they were
scaled according to oxygen. Nitrogen was assumed to be partly primary
and partly secondary and it was scaled as [N/O] = 0.08 + [O/H]
\citep{denicolo02} which is in accordance with the observed behaviour
of N/O with O/H \citep{henry00}. The Orion nebula dust properties were
assumed and, as a starting value, Orion nebula abundances were
used. With these input parameters a grid of models was calculated for
various values of the ionisation parameter and metallicity, and the
results are shown in Fig.~\ref{diagram}. Further models with higher
and smaller densities (5000 cm$^{-3}$/10 cm$^{-3}$), different shapes
for the ionising sources (blackbody continuum/Kurucz), different
filling factor values, up to = 1, and cloud sizes and geometry
were calculated. In general, the overall changes with respect to the
grid of models shown in Fig.~\ref{diagram} were small, with
differences $\la$ 0.15 dex in the model metallicities.




Fig.~\ref{diagram} shows that, for most of the area occupied by the
H\,\textsc{ii} regions, the modelling predicts single valued
metallicity and ionisation parameter diagnoses. Only at the highest
metallicities and lower ionisation parameter, some folding is
present. It is interesting to notice that our modelling gives single
valued results down to $\sim$ log(H$\alpha$/[N\,\textsc{ii}]) = 0,
while e.g. \citet{melbourne02} have shown that at
log(H$\alpha$/[N\,\textsc{ii}]) $<$ 0.45 the nitrogen line ratio alone
ceases to be a good metallicity indicator. The model metallicities
agree well with the average values adopted for the different galaxies
(Table \ref{table1}), except for the highest abundances where
\textsc{CLOUDY} predicts lower abundance values. In particular, the
O/H abundance of M31 is quoted as 12 + log(O/H) = 9.0 by 
\citet{van_den_bergh00} but
most of the H\,\textsc{ii} regions of M31 occupy a region between 8.4 and
 8.8 in Fig.~\ref{diagram}. This suggests that the
O/H abundance adopted by \citet{van_den_bergh00} for M31 could be
significantly overestimated, as it was the case for other spiral
galaxies studied by \citet{pilyugin07}. These authors found a maximum
metallicity of 12 + log(O/H) $=$ 8.87 for the H\,\textsc{ii}
regions located at the bulges of the most metal-rich spiral galaxies
\citep[see also][]{bresolin04,pilyugin06}.

We have compared the O/H abundances estimated through the S2N2 ratio 
to the abundances calculated by \citet{magrini07}
for 14 H\,\textsc{ii} regions in M33 (using the empirical ICF method)
and by \citet{magrini05} for H\,\textsc{ii} regions in Sextans A (8
regions) and Sextans B (7 regions) using \textsc{CLOUDY}
modelling. The median differences in 12 + log(O/H) between the two sets
of measurements were 0.24 for Sex A, 0.13 for SexB, and 0.20 for M33,
and they do not show any systematics. Considering the quoted errors in the
direct O/H abundance determination \citep[$\geq$ 0.1 for Sextans A and B and
0.04-0.25 for M33;][]{magrini05,magrini07}, these discrepancies
seem acceptable.

We have explored the use of the S2N2 indicator for high redshift HII
galaxies in the sample of \citet{shapley04}. These authors estimated
the O/H abundance of several galaxies at $z=$2.1-2.5; the objects are
therefore located at the so called ``redshift-desert'' (1.4 $<$ $z$ $<$ 2.5)
where there were virtually no previous determinations of the
H\,\textsc{ii} region metallicities. These are crucial objects indeed
since
that epoch may host the production of a large fraction of the heavy
elements present in the local universe. At these redshifts the lines
needed for the S2N2 diagram are observable in the near-infrared
($K$-band).
We have derived the O/H abundances for the two galaxies in the
\citet{shapley04} sample for which both the nitrogen and sulphur lines
were measured with good accuracy: Q1623-MD~66 ($z$ = 2.1) N2 = 0.77; S2 =
0.61, and Q1623-BX~453 ($z$ = 2.2) N2 = 0.52; S2 = 0.75
(\citet{shapley04}, and private communication by Dr. Shapley). We get
12 + log(O/H) = 8.3 (MD~66) and 8.7 (BX~453), to be compared to the
corresponding measurements by \citet{shapley04}, based on
\citet{pettini04} and \citet{denicolo02} N2 calibrations, giving 8.5
and 8.6, respectively, for MD~66, and 8.6 and 8.7, respectively, for
BX~453.
It is clear that the three sets of measurements differ, albeit the
main results of the \citet{shapley04} study hold: both galaxies show a
high metallicity (in the case of BX~453, comparable to the most
metallic nearby spirals) with MD~66 being oxygen poorer than
BX~453. In the case of MD~66, our O/H abundance is 0.2-0.3 dex lower
than in \citet{shapley04}. It should be noted that our modelling would
permit both the \citet{pettini04} and \citet{denicolo02} O/H values
for this galaxy if only the nitrogen line ratio would be
considered. The sulphur line ratio helps fixing better the metallicity
(see Fig.~\ref{diagram}). On the other hand, as mentioned by
\citet{shapley04}, their O/H measurement for BX~453 may be affected by
the saturation of [N\,\textsc{ii}]/H$\alpha$ - O/H diagnostic at high
metallicities. Our modelling shows saturation only at N2 $\la$ 0, and
should not thus affect the derived O/H abundances for the galaxies in
question here.

One important issue with all O/H measurements in unresolved distant
galaxies is, as \citet{shapley04} note, that the diffuse interstellar
medium in each galaxy (DIG) can make a significant contribution to the
observed spectra. According to \citet{moustakas06}, the presence of a
strong DIG decreases both the N2 and S2 ratios and would thus lead to
overestimating the metallicity.






\begin{figure*}
   \centering
   \includegraphics[width=0.5\linewidth,angle=-90]{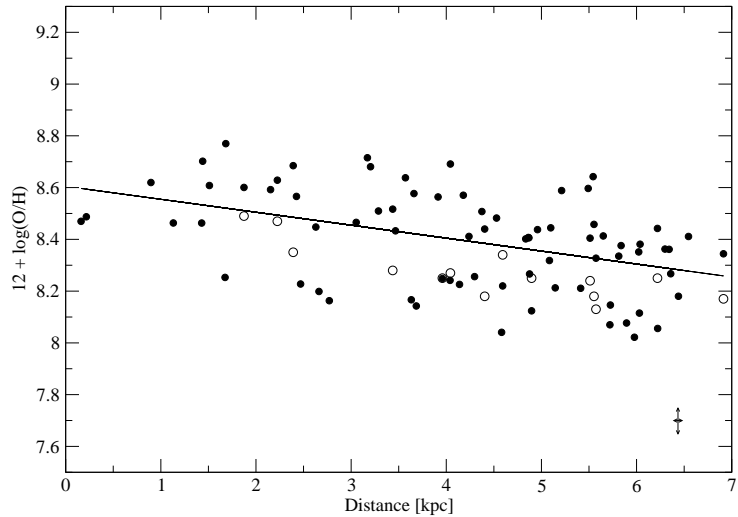}     
\caption{The O/H abundance versus the galacto-centric distance for 76 H\,{\sc ii} regions of M33. {\it {Filled circles}}: abundances derived from our modelling. The solid line is the least square fit to these data. {\it {Open circles}}: abundances for 14 H\,{\sc ii} regions measured by \citet{magrini07} using the empirical method. } 
\label{gradient}
    \end{figure*}

\section{The O/H gradient in M33}

Many galaxies show radial abundance gradients, with the metallicity
decreasing from the centre towards the edge of the galaxy. One of the
best galaxies to study the abundance gradient is M33, for its
closeness, wealth of H\,\textsc{ii} regions, and adequate orientation
(almost face-on). \citet{vilchez88} derived a global abundance
gradient of -0.12 dex kpc$^{-1}$ for M33. This measurement was based
on their own (direct) abundance measurements in seven H\,\textsc{ii}
regions, plus H\,\textsc{ii} region and SNR abundance measurements by
other authors \citep[see][ and references therein]{vilchez88}. Using
H\,\textsc{ii} region data compiled from literature \citet{garnett97}
derived a similar O/H gradient of -0.11 dex
kpc$^{-1}$. \citet{monteverde97} measured for the stellar (B
super-giant) oxygen abundance gradient a value of -0.16 $\pm$ 0.06 dex
kpc$^{-1}$. However, a more recent estimation of \citet{urbaneja05}
for a sample of 11 B-type stars yielded a smaller value of -0.06 $\pm$
0.02 dex kpc$^{-1}$. \citet{magrini04} measured the chemical
abundances for 11 planetary nebulae in M33 and found an O/H gradient
of -0.14 dex kpc$^{-1}$, i.e. in agreement with the H\,\textsc{ii}
region gradient. Recent abundance measurements of six H\,\textsc{ii}
regions by \citet{crockett06} show a gradient -0.012 $\pm$ 0.011 dex
kpc$^{-1}$, significantly shallower than the previous
estimations. However, based on a sample of 5 Beat Cepheids,
\citet{beaulieu06} inferred again a steeper gradient, -0.16 dex
kpc$^{-1}$. Finally, \citet{perez-montero05} have studied the M33 O/H
gradient based on data of $\sim$ 20 H\,\textsc{ii} regions (using both
the direct method and the empirical calibrator S$_{23}$/O$_{23}$) and
early type stars. Albeit they do not provide a numeric value for the
gradient, they conclude that the agreement between empirical and
direct abundance gradients for M33 is good.

Our data sample includes 76 H\,\textsc{ii} regions in M33, most of
which are new measurements by \cite{magrini07}. O/H abundances for all
of them were individually determined with the modelling described
above. As for their galacto-centric distances, \cite{magrini07}
provide coordinates for the centres of the H\,\textsc{ii} regions in
their sample, while for the rest of the regions the coordinates are
given in \citet{kwitter81}.
We adopt an inclination angle of 53$\degr$, position angle of major
axis of 23$\degr$ and a distance of 840 kpc. 
The H\,{\sc ii} regions cover the galacto-centric distances from 0.16 kpc
 up to 6.91 kpc thus approaching the galactic edge until 0.7R$_{25}$ 
(HyperLeda Extragalactic Database). 
 Fig.~\ref{gradient} shows
the resultant oxygen abundances as a function of the galacto-centric
distances of the corresponding H\,\textsc{ii} regions. A linear fit
gives a value of 12 + log(O/H) = -0.05 ($\pm$0.01) dex kpc$^{-1}$ +
8.60 ($\pm$0.05). 

The errors in the metallicity estimations consist of uncertainties in the
modelling plus errors in the measured line ratios. Due to the large
number of parameters involved, realistic errors in the modelling are
difficult to estimate.  
As mentioned in section \ref{sec3}, changing the model input parameters to their extreme values can lead into differences of 0.15 dex in the Oxygen abundance. However, for a given model, an error of 0.04 dex in the line
 ratios (most H\,\textsc{ii} regions in our sample have errors
smaller than this)
causes an error typically of $\sim$0.1 in 12 + log(O/H). On the other
hand, errors in the galacto-centric distances are small, around 0.1
kpc \citep{magrini07}.  These error-bars are shown in the right-down
corner of Fig.~\ref{gradient}.

The mean deviation of our data points from the best fit line in Fig.~\ref{gradient} is 0.17 dex. The error bar shown in Fig.~\ref{gradient} plus the possible uncertainty caused by the modelling (0.15 dex, see above) are then enough to explain this scatter. However, there may also exist a real scatter in the abundances. This could be caused by localised gas infall and ineffective mixing or spatially asymmetric chemical enrichment in the galaxy \citep[for theoretical discussion, see][]{roy95}. These points would certainly be interesting for a detailed future study.

Finally, the oxygen abundance gradient calculated here can be compared
to the corresponding gradient by \cite{magrini07}, based on 14
H\,\textsc{ii} regions with directly measured (ICF) abundances. These 
measurements are presented in Fig.3 as open circles. Their
gradient is -0.06 ($\pm$0.01) dex kpc$^{-1}$ + 8.53 ($\pm$0.05),
i.e. in excellent agreement with our determination in this
paper. However, our estimation of the metallicity at the centre of M33
is 0.10 higher than the corresponding value by \cite{magrini07}.

\section{Conclusions}

We have shown that the H$\alpha$/S[\,\textsc{ii}],
H$\alpha$/N[\,\textsc{ii}] (S2N2) diagram is a powerful metallicity
and ionisation parameter indicator for extragalactic H\,\textsc{ii}
regions. A clear advantage of the S2N2 diagram is that the wavelengths
of the lines used are close to each other, and thus are not affected
by uncertainties in the reddening correction and in the calibration of
the instrumental spectral response.  The diagram gives unambiguous
results over a wide metallicity range. Furthermore, the diagram is
especially useful for studies of large samples of new emission line
objects of unknown nature, as it separates very efficiently
H\,\textsc{ii} regions, PNe, and SNRs from each others, while enabling
to estimate their O/H abundances. On the other hand, the
[S\,\textsc{ii}] lines are usually rather weak and may thus not be
detectable in all H\,\textsc{ii} regions, whereas the nitrogen lines
are possibly influenced by the galaxy's star forming history through
the evolution of the N/O abundance ratio.

At low and intermediate metallicities, the diagram is single valued
with respect to metallicity; at higher metallicities some folding is
present. At the highest metallicities there is also a discrepancy
between the O/H abundances predicted by the model and the empirical
H\,{\sc ii} region abundances. However, these can be reconciled if the
latter would be significantly lower
than previously believed, as proposed by \citet{bresolin04} and
\citet{pilyugin07}.

Based on a sample of 76 H\,\textsc{ii} region measurements \citep[12
from the literature and 64 new measurements by][]{magrini07} we have
revisited the abundance gradient of M33. We derive for the gradient a
value of -0.05 $\pm$ 0.01 dex kpc$^{-1}$ in agreement with the recent
direct measurements by \cite{magrini07}.


\section*{Acknowledgements}

The work of K. Viironen was supported by an undergraduate grant
from the University of Oulu and a grant from the foundation of Magnus 
Ehrnrooth, Finland, plus a PhD grant from the Instituto de 
Astrof{\' {i}}sica de
Canarias, Spain. Continuous funding form the Spanish Ministerio de
Educaci\'on y Ciencia (grant AYA-02-00883) is sincerely
acknowledged. G. D-I. acknowledges support from Mexican CONACYT projects 42611-F and 50359-F. We are grateful to Dr. Alice Shapley for kindly
providing us with the sulphur line data for the galaxies Q1623-MD~66
and Q1623-BX~453.

\bibliographystyle{mn2e}
\bibliography{ref}

\appendix

\section{N2 and S2 data for H\,{\sc ii} regions in Local Group galaxies}

\begin{table*}
\begin{minipage}{150mm}
\caption{The H\,{\sc ii} region sample in Local Group galaxies. The errors in the line ratios are given in brackets.}
\label{table2}
\begin{tabular}{cccccccc}
\hline
\hline
region & log(H$\alpha$/[N\,{\sc ii}]) & log(H$\alpha$/[S\,{\sc ii}]) & reference & region & log(H$\alpha$/[N\,{\sc ii}]) & log(H$\alpha$/[S\,{\sc ii}]) & reference \\
\hline
\hline
\multicolumn{8}{c}{M31}\\
\hline
K59 & 0.74 (0.05) & 0.66 (0.04) & 1 &
K68 & 0.39 (0.03) & 0.85 (0.03) & 1 \\
K70 & 0.51 (0.03) & 0.49 (0.03) & 1 &
K76 & 0.61 (0.05) & 0.49 (0.03) & 1 \\
K78 & 0.57 (0.05) & 0.52 (0.03) & 1 &
K81 & 0.59 (0.05) & 0.46 (0.02) & 1 \\
K87 & 0.47 (0.05) & 0.46 (0.04) & 1 &
K91 & 0.47 (0.03) & 0.57 (0.03) & 1 \\
K92 & 0.46 (0.03) & 0.47 (0.03) & 1 &
K132 & 0.45 (0.02) & 0.68 (0.02) & 1 \\
K145 & 0.42 (0.06) & 0.46 (0.05) & 1 &
K244 & 0.40 (0.03) & 0.57 (0.03) & 1 \\
K250 & 0.45 (0.06) & 0.66 (0.06) & 1 &
K310 & 0.55 (0.05) & 0.51 (0.04) & 1 \\
K314 & 0.42 (0.03) & 0.42 (0.02) & 1 &
K315 & 0.38 (0.06) & 0.60 (0.07) & 1 \\
K330 & 0.46 (0.03) & 0.548 (0.015) & 1 &
K343 & 0.39 (0.03) & 0.66 (0.02) & 1 \\
K353 & 0.24 (0.04) & 0.28 (0.03) & 1 &
K403 & 0.89 (0.07) & 0.52 (0.03) & 1 \\
K414 & 0.49 (0.03) & 0.55 (0.02) & 1 &
K434 & 0.50 (0.06) & 0.40 (0.03) & 1 \\
K442 & 0.61 (0.05) & 0.92 (0.04) & 1 &
K446 & 0.53 (0.04) & 0.54 (0.03) & 1 \\
K447 & 0.41 (0.03) & 0.62 (0.04) & 1 &
K461 & 0.18 (0.03) & 0.37 (0.03) & 1 \\
K480 & 0.45 (0.06) & 0.59 (0.07) & 1 &
K496 & 0.41 (0.03) & 0.54 (0.03) & 1 \\
K525 & 0.43 (0.12) & 0.46 (0.06) & 1 &
K526 & 0.52 (0.03) & 0.72 (0.02) & 1 \\
K536 & 0.45 (0.05) & 0.42 (0.02) & 1 &
K703 & 0.49 (0.04) & 0.43 (0.02) & 1 \\
K722 & 0.46 (0.05) & 0.74 (0.07) & 1 &
K787 & 0.62 (0.04) & 0.46 (0.02) & 1 \\
K838 & 0.53 (0.07) & 0.46 (0.05) & 1 &
K851 & 0.43 (0.12) & 0.15 (0.06) & 1 \\
K877 & 0.42 (0.05) & 0.57 (0.03) & 1 &
K927 & 0.45 (0.02) & 0.62 (0.02) & 1 \\
K931 & 0.46 (0.05) & 0.36 (0.04) & 1 &
K932 & 0.44 (0.06) & 0.52 (0.04) & 1 \\
BA75 & 0.50 (0.04) & 0.33 (0.03) & 2 &
BA423 & 0.50 (0.03) & 0.62 (0.02) & 2 \\
BA289 & 0.80 (0.03) & 1.05 (0.05) & 2 \\
\hline
\multicolumn{8}{c}{M33}\\
\hline
CC93 & 0.54 (0.02) & 0.51 (0.02) & 3 &
IC142 & 0.69 (0.03) & 0.77 (0.02) & 3 \\
NGC595 & 0.816 (0.012) & 1.06 (0.012) & 3 &
MA2 & 1.03 (0.03) & 1.06 (0.04) & 3 \\
NGC604 & 0.902 (0.003) & 0.887 (0.002) & 3 &
NGC588 & 1.55 (0.03) & 1.25 (0.04) & 3 \\
MA11 & 0.82 (0.04) & 0.90 (0.04) & 4 &
MA3 & 1.22 (0.04) & 1.33 (0.04) & 4 \\
IC131 & 1.23 (0.04) & 1.21 (0.04) & 4 &
MA9a & 0.92 (0.04) & 0.84 (0.04) & 4 \\
IC133 & 1.39 (0.04) & 1.41 (0.04) & 4 &
IC132 & 1.6 (0.04) & 1.39 (0.04) & 4 \\
BCLMP~275  &  0.94 (0.03)  &  0.75 (0.02)  & 5 &
BCLMP~272  &  0.87 (0.03)  &  0.44 (0.02)  & 5 \\
BCLMP~273  &  0.89 (0.07)  &  0.54 (0.04)  & 5 &
BCLMP~266  &  0.97 (0.03)  &  0.69 (0.02)  & 5 \\
BCLMP~263  &  0.83 (0.03)  &  0.507 (0.015)  & 5 &
LGC-HII-2  &  0.77 (0.02)  &  0.715 (0.013)  & 5 \\
BCLMP~238  &  0.98 (0.03)  &  0.91 (0.04)  & 5 &
LGC-HII-3  &  0.945 (0.011)  &  0.851 (0.010)  & 5 \\
BCLMP~240  &  0.95 (0.03)  &  0.73 (0.03)  & 5 &
BCLMP~289  &  0.93 (0.04)  &  0.63 (0.03)  & 5 \\
BCLMP~218  &  0.934 (0.009)  &  0.731 (0.006)  & 5 &
BCLMP~258  &  1.09 (0.02)  &  0.73 (0.02)  & 5 \\
BCLMP~288  &  1.02 (0.02)  &  1.03 (0.02)  & 5 &
MCM2000~Em~24  &  0.917 (0.012)  &  0.863 (0.015)  & 5 \\
BCLMP~220  &  1.21 (0.02)  &  0.916 (0.012)  & 5 &
CPSDP~194  &  1.132 (0.005)  &  1.230 (0.006)  & 5 \\
BCLMP~256  &  1.08 (0.03)  &  1.04 (0.02)  & 5 &
VGHC~2-22  &  1.031 (0.015)  &  0.969 (0.013)  & 5 \\
BCLMP~618  &  0.891 (0.008)  &  0.825 (0.010)  & 5 &
CPSDP~103  &  0.89 (0.03)  &  0.66 (0.02)  & 5 \\
BCLMP~638N  &  1.36 (0.02)  &  1.18 (0.02)  & 5 &
BCLMP~626  &  1.02 (0.02)  &  0.866 (0.014)  & 5 \\
CPSDP~107  &  1.21 (0.04)  &  0.94 (0.03)  & 5 &
LGC-HII-4  &  0.83 (0.02)  &  0.89 (0.03)  & 5 \\
BCLMP~45  &  1.112 (0.007)  &  1.135 (0.007)  & 5 &
BCLMP~207  &  0.52 (0.02)  &  0.117 (0.013)  & 5 \\
BCLMP~640  &  1.18 (0.06)  &  0.83 (0.03)  & 5 &
BCLMP~55  &  0.94 (0.03)  &  0.74 (0.02)  & 5 \\
CDL2004~52  &  0.617 (0.006)  &  0.582 (0.008)  & 5 &
CPSDP~221  &  0.786 (0.009)  &  0.885 (0.015)  & 5 \\
LGC-HII-6  &  0.859 (0.007)  &  0.99 (0.02)  & 5 &
BCLMP~637  &  1.17 (0.02)  &  0.93 (0.02)  & 5 \\
BCLMP~93  &  0.633 (0.006)  &  0.617 (0.008)  & 5 &
IC~142  &  0.717 (0.014)  &  0.68 (0.02)  & 5 \\
BCLMP~759  &  0.93 (0.03)  &  0.74 (0.02)  & 5 &
BCLMP~4  &  0.829 (0.006)  &  0.985 (0.010)  & 5 \\
GDK99~128  &  0.836 (0.003)  &  0.922 (0.005)  & 5 &
CPSDP~316  &  0.91 (0.03)  &  0.82 (0.03)  & 5 \\
BCLMP~670  &  0.687 (0.012)  &  0.693 (0.014)  & 5 &
LGC-HII-7  &  1.09 (0.03)  &  0.96 (0.07)  & 5 \\
VGHC-2-84  &  0.912 (0.005)  &  1.040 (0.009)  & 5 &
BCLMP~709  &  0.832 (0.010)  &  0.867 (0.014)  & 5 \\
CPSDP~323  &  0.84 (0.02)  &  0.76 (0.03)  & 5 &
CPSDP~325  &  0.67 (0.05)  &  0.46 (0.06)  & 5 \\
BCLMP~721  &  0.93 (0.03)  &  0.69 (0.02)  & 5 &
BCLMP~682  &  0.75 (0.03)  &  0.70 (0.02)  & 5 \\
BCLMP~715  &  0.78 (0.02)  &  0.50 (0.02)  & 5 &
CPSDP~189  &  0.80 (0.06)  &  0.59 (0.05)  & 5 \\
BCLMP~659  &  0.95 (0.03)  &  0.91 (0.03)  & 5 &
LGC-HII-8  &  0.89 (0.04)  &  0.62 (0.02)  & 5 \\
BCLMP~680  &  0.70 (0.02)  &  0.64 (0.02)  & 5 &
BCLMP~717  &  0.950 (0.010)  &  0.880 (0.010)  & 5 \\
LGC-HII-10  &  0.87 (0.03)  &  0.78 (0.02)  & 5 &
BCLMP~749  &  0.927 (0.012)  &  0.96 (0.02)  & 5 \\
BCLMP~649  &  1.04 (0.03)  &  0.90 (0.03)  & 5 &
BCLMP~705  &  0.75 (0.03)  &  0.81 (0.06)  & 5 \\
LGC-HII-11  &  1.16 (0.02)  &  1.00 (0.02)  & 5 &
MCM2000~Em~59  &  0.71 (0.02)  &  0.63 (0.02)  & 5 \\
BCLMP~757  &  0.99 (0.02)  &  0.896 (0.012)  & 5 &
BCLMP~657  &  1.02 (0.04)  &  0.86 (0.04)  & 5 \\
LGC-HII-12  &  0.91 (0.02)  &  0.77 (0.02)  & 5 &
BCLMP~754  &  1.01 (0.02)  &  0.67 (0.02)  & 5 \\
LGC-HII-13  &  1.24 (0.05)  &  1.11 (0.06)  & 5 &
BCLMP~756  &  0.995 (0.014)  &  0.88 (0.02)  & 5 \\
\hline
\end{tabular}
\end{minipage}
\end{table*}
\begin{table*}
\begin{minipage}{150mm}
\contcaption{}
\begin{tabular}{cccccccc}
\hline
\hline
region & log(H$\alpha$/[N\,{\sc ii}]) & log(H$\alpha$/[S\,{\sc ii}]) & reference & region & log(H$\alpha$/[N\,{\sc ii}]) & log(H$\alpha$/[S\,{\sc ii}]) & reference \\
\hline
\hline
\multicolumn{8}{c}{LMC}\\
\hline
N160A1 & 1.18 (0.04) & 1.10 (0.04) & 6 &
N160A2 & 1.19 (0.05) & 1.13 (0.05) & 6 \\
N159-5 & 1.13 (0.04) & 0.95 (0.03) & 6 &
N157B & 1.11 (0.05) & 0.73 (0.03) & 6 \\
N11A & 1.23 (0.04) & 1.14 (0.05) & 6 &
N83B & 1.10 (0.03) & 1.04 (0.03) & 6 \\
N79A & 1.08 (0.03) & 1.02 (0.03) & 6 &
N4A & 1.33 (0.03) & 1.25 (0.03) & 6 \\
30Dor1 & 1.74 (0.04) & 1.62 (0.03) & 7 &
30Dor2 & 1.78 (0.04) & 1.49 (0.03) & 7 \\
30Dor3 & 1.56 (0.03) & 1.53 (0.03) & 7 &
30Dor4 & 1.35 (0.03) & 1.26 (0.03) & 7 \\
NGC2079 & 1.29 (0.12) & 1.12 (0.11) & 7 &
IC2111 & 1.11 (0.07) & 1.06 (0.09) & 7 \\
N157 & 1.53062 &  1.43861 & 8 &
N159 & 1.28913 &  1.12494 & 8 \\
N8 & 1.39 (0.03) & 1.20 (0.03)  & 9 &
N11B & 1.28 & 1.11 & 10 \\
\hline
\multicolumn{8}{c}{IC 10}\\
\hline
HL106a & 0.92 (0.06) & 0.62 (0.03) & 11 &
HL106b & 1.16 (0.06) & 1.03 (0.06) & 11 \\
HL111b & 0.96 (0.03) & 0.72 (0.02) & 11 &
HL111c(WR) & 1.31 (0.09) & 1.12 (0.02) & 11 \\
HL111e & 1.09 (0.04) & 0.81 (0.02) & 11 &
HL30 & 1.07 (0.08) & 0.84 (0.06) & 11 \\
HL45 & 1.48 (0.05) & 1.172 (0.011) & 11 &
HL111d+e & 1.25 (0.07) & 1.11 (0.08) & 11 \\
HL77 & 1.2 (0.1) & 1.03 (0.09) & 11 &
IC10(1) & 1.42 (0.03) & 1.21 (0.03) & 12 \\
IC10(2) & 1.39 (0.03) & 1.26 (0.03) & 12 \\
\hline
\multicolumn{8}{c}{NGC6822}\\
\hline
Ho11 & 1.27 (0.15) & 1.03 (0.15) & 13 &
HuX & 1.74 (0.10) & 1.30 (0.07) & 13 \\
HuV & 1.71 (0.07) & 1.38 (0.03) & 13 &
Ho14 & 0.86 (0.10) & 0.80 (0.11) & 13 \\
Ho13 & 1.28 (0.08) & 0.76 (0.06) & 13 &
Ho15 & 1.50 (0.15) & 1.20 (0.08) & 13 \\
HuV & 1.75 (0.02) & 1.386 (0.013) & 14 &
HuX & 1.60 (0.03) & 1.21 (0.02) & 14 \\
HK16 & 0.91 (0.13) & 0.45 (0.06) & 15 &
Ho12 & 1.03 (0.09) & 0.27 (0.03) & 15 \\
HuV & 2.31 (0.04) & 1.45 (0.03) & 15 &
Kalfa & 2.09 (0.06) & 1.49 (0.02) & 15 \\
KD28e & 1.61 (0.07) & 1.11 (0.08) & 15 \\
\hline
\multicolumn{8}{c}{SMC}\\
\hline
N88A & 1.73 (0.03) & 1.47 (0.03) & 6 &
N66 & 1.67 (0.05) & 1.22 (0.04) & 6 \\
N12A & 1.48 (0.06) & 1.08 (0.06) & 9 &
N12B & 1.66 (0.06) & 1.06 (0.06) & 9 \\
N22 & 1.57 (0.06) & 1.23 (0.06) & 9 &
N25 & 1.37 (0.06) & 1.02 (0.06) & 9 \\
N76SW & 1.49 (0.06) & 0.94 (0.06) & 9 &
N83 & 1.47 (0.06) & 1.17 (0.06) & 9 \\
N90 & 1.53 (0.06) & 1.24 (0.06) & 9 &
NGC346 & 1.873 (0.005) & 1.365 (0.003) & 16 \\
N81 & 1.66 (0.12) & 1.4 (0.1) & 17 \\
\hline
\multicolumn{8}{c}{WLM}\\
\hline
WLM1 & 1.43 (0.09) & 0.85 (0.04) & 18 \\
\hline
\multicolumn{8}{c}{SextansA}\\
\hline
HSK2 & 1.60 (0.12) & 0.94 (0.04) & 19&
HSK3 & 1.57 (0.14) & 1.06 (0.06) & 19\\
HSK7 & 1.54 (0.10) & 1.05 (0.04) & 19&
HSK9 & 1.41 (0.08) & 0.94 (0.04) & 19\\
HSK16 & 1.59 (0.11) & 0.89 (0.03) & 19&
HSK19 & 1.47 (0.13) & 0.85 (0.03) & 19\\
HSK23 & 1.40 (0.07) & 0.96 (0.03) & 19&
HSK24 & 1.45 (0.13) & 0.79 (0.04) & 19  \\
\hline
\multicolumn{8}{c}{SextansB}\\
\hline
SextB1 & 1.38 (0.09) &	0.91 (0.03) & 17 &
SextB3 & 1.44 (0.14) &	0.85 (0.04) & 17 \\
SHK2 & 1.318 (0.014) & 0.804 (0.013) & 19 &
SHK3 & 1.15 (0.07) & 0.54 (0.02) & 19 \\
SHK5 & 1.60 (0.05) & 0.95 (0.03) & 19 &
SHK6 & 1.10 (0.11) & 0.44 (0.04) & 19 \\
SHK7 & 1.13 (0.04) & 0.77 (0.03) & 19 &
SHK10 & 1.20 (0.02) & 0.812 (0.011) & 19 \\
HIIH & 1.50 (0.04) & 0.92 (0.02) & 19 &
SextB &	1.62 (0.12) &	0.84 (0.11) &  20 \\
\hline
\multicolumn{8}{c}{LeoA}\\
\hline
-101-052 & 1.89 (0.10) & 1.18 (0.04) & 21&
-091-048 & 1.58 (0.08) & 1.10 (0.05) & 21\\
+069-018 & 1.52 (0.07) & 0.98 (0.05) & 21&
+112-020 & 1.62 (0.04) & 1.02 (0.03) & 21\\
\hline
\multicolumn{8}{c}{Gr8}\\
\hline
-019-019 & 1.52 (0.03) & 1.01 (0.03) & 21&
-013-032 & 1.30 (0.05) & 0.76 (0.04) & 21\\
-012-022 & 1.53 (0.03) & 0.97 (0.03) & 21&
-002-012 & 1.34 (0.03) & 0.74 (0.03) & 21\\
+001-008 & 1.52 (0.04) & 0.82 (0.03) & 21&
+001+027 & 1.55 (0.05) & 1.01 (0.04) & 21\\
+004-006 & 1.37 (0.03) & 0.81 (0.02) & 21&
+008-011 & 1.46 (0.03) & 0.99 (0.02) & 21\\
\hline
\end{tabular}
\end{minipage}
\end{table*}
\begin{table*}
\begin{minipage}{150mm}
\contcaption{}
\begin{tabular}{cccccccc}
\hline
\hline
region & log(H$\alpha$/[N\,{\sc ii}]) & log(H$\alpha$/[S\,{\sc ii}]) & reference & region & log(H$\alpha$/[N\,{\sc ii}]) & log(H$\alpha$/[S\,{\sc ii}]) & reference \\
\hline
\hline
\multicolumn{8}{c}{IC 1613}\\
\hline
IC1613.13 & 1.32 (0.11) & 0.87 (0.07) & 22 & & & & \\
\hline
\multicolumn{8}{c}{NGC 3109}\\
\hline
NGC3109.6.6 & 1.24 (0.12) & 0.92 (0.08) & 22 & & & \\
\hline
\end{tabular}

\medskip
References: (1) \citet{galarza99}\footnote{\citet{galarza99} present line measurements with various apertures, covering different parts of the H\,\textsc{ii} region or the whole complex in their sample. We have used the fluxes corresponding to the whole complex, when available, or to the brightest zone in each region.}, (2) \citet{blair82}, (3) \citet{vilchez88}, (4) \citet{kwitter81}, (5) \citet{magrini07}, (6) \citet{vermeij02}, (7) \citet{peimbert74}, (8) \citet{dufour82},(9) \citet{dufour77b}, (10) \citet{tsamis03}, (11) \citet{richer01}, (12) \citet{lequeux79}, (13) \citet{pagel80}, (14) \citet{peimbert05}, (15) \citet{lee06}, (16) \citet{peimbert00}, (17) \citet{hunter99}, (18) \citet{skillman89}, (19) \citet{magrini05}, (20) \citet{moles90}, (21) \citet{van_zee06}, (22) \citet{lee03}
\end{minipage}
\end{table*}


\label{lastpage}

\end{document}